\documentclass[prl,aps,preprint]{revtex4-1}
\usepackage{graphicx}

\begin{document}
\title{Mesoscopic superconductivity and high spin polarization coexisting at metallic point contacts on the Weyl semimetal TaAs}

\author{Leena Aggarwal$^1$}
\author{Sirshendu Gayen$^1$}
\author{Shekhar Das$^1$}
\author{Ritesh Kumar$^1$}
\author{Vicky S\"{u}{\ss}$^2$}
\author{Chandra Shekhar$^2$}
\author{Claudia Felser$^2$}
\author{Goutam Sheet$^1$}
\email {goutam@iisermohali.ac.in}
\affiliation{$^1$Department of Physical Sciences,
Indian Institute of Science Education and Research Mohali,
Sector 81, S. A. S. Nagar, Manauli, PO: 140306, India}
\affiliation{$^2$Max Planck Institute for Chemical Physics of Solids, N\"{o}thnitzer Stra{\ss}e 40, 01187 Dresden, Germany.}
\date{\today}

\begin{abstract}
\textbf{A Weyl semimetal\cite{WSM1, WSM2, WSM3, WSM4, WSM5, WSM6, WSM7, WSM8, WSM9, WSM10,WSM11} is a topologically non-trivial phase of matter that hosts mass-less Weyl fermions,\cite{WF1} the particles that remained elusive for more than 80 years since their theoretical discovery. The Weyl semimetals exhibit unique transport and magneto-transport properties\cite{Prop} and remarkably high surface spin polarization.\cite{WSM_SP} Here we show that a unique mesoscopic superconducting phase with a critical temperature up to 7 K can be realized by forming metallic point contacts with silver (Ag) on single crystals of TaAs, while neither Ag nor TaAs are superconductors. The Andreev reflection spectra obtained from such point contacts are fitted well within a modified Blonder-Tinkham-Klapwijk (BTK) model\cite{BTK,Mazin,Soulen,Anshu} with a superconducting energy gap up to 1.2 meV. The analysis within this model also reveals high transport spin polarization up to 60\% indicating a spin polarized supercurrent flowing through the point contacts on TaAs. Such point contacts also show a large anisotropic magnetoresistance (AMR) originating from the spin polarized current.\cite{AMR, AMR1} Therefore, apart from the discovery of a novel mesoscopic superconducting phase and it's coexistence with a large spin polarization, our results also show that the point contacts on Weyl semimetals are potentially important for applications in spintronics.}      
\end{abstract}

\maketitle

The discovery of Weyl semimetals\cite{WSM3, WSM4, WSM13} facilitated the realization of Weyl fermions in condensed matter systems after more than 80 years of their theoretical discovery\cite{WSM12}. In quantum field theory, the Weyl fermions were first shown by Hermann Weyl to emerge as solutions to the relativistic Dirac equation.\cite{WSM3,WSM10} However, until the discovery of TaAs as a Weyl semimetal \cite{WSM3,WSM4,WSM6,WSM7,WSM8,WSM9,WSM10},WSM, such exotic particles remained elusive in nature. The Weyl semimetals are topologically non-trivial and are known to demonstrate exotic quantum phenomena and unique surface states.\cite{Prop} The band structure of a Weyl semimetal involves Weyl nodes, which can be imagined as a monopole or an antimonopole of the Berry curvature in the momentum space.\cite{WSM7,WSM15} Each of the Weyl nodes are associated with a quantized chiral charge and the Weyl nodes are connected with each other only through the boundary of the crystals via Fermi arcs, the characterizing topological surface states in a Weyl semimetal.\cite{WSM3,WSM6,WSM7,WSM10,WSM16} Recently it was shown that the Fermi arcs in the Weyl semimetal TaAs are highly spin polarized which lie in a completely 2D plane on the surface of the crystal.\cite{WSM_SP} As a consequence of such exotic topological properties, the Weyl semimetals are believed to host even richer set of physical phenomena that must be explored for better understanding of quantum mechanics and for potential device applications. In this article, for the first time, we show the emergence of a unique superconducting phase coexisting with the highly spin-polarized surface states at mesoscopic point contacts between elemental normal metals and high quality single crystals of the Weyl semimetal TaAs. The critical temperature ($T_c$) of such a superconducting point contacts found to be 7 K.

Since a tip-induced superconducting (TISC) phase emerges only under point contacts, the traditional bulk characterization tools for characterizing superconducting phases fail to detect TISC. However, by performing point contact spectroscopy of such point contacts it is possible to explore different regimes of mesoscopic transport where the expected spectral features for superconducting point contacts are well understood. Such experiments can reveal mesoscopic superconducting phases like a TISC that emerge only under point contacts on exotic materials.\cite{WSM19,WSM20} Furthermore, by driving the point contacts to the ballistic regime of transport, it is also possible to extract spectroscopic information of the superconducting phase.\cite{WSM17}

In Figure 1 (a) we show the schematic of point contacts made on single crystals of TaAs. In Figure 1(b) we show a point contact spectrum between silver (Ag) and a single crystal of TaAs. The spectrum shows two conductance dips and a zero-bias conductance peak. Such spectra are usually seen for superconducting point contacts in the thermal regime of transport where critical current dominated effects give rise to the conductance dips.\cite{WSM17, WSM18} For certain superconducting point contacts in the thermal limit, when a single physical point contact comprises of multiple electrical micro-contacts, multiple critical current driven dips can be expected. One such spectrum is shown in Figure 1(c). The magnetic field dependence of this spectrum is shown in Figure 1(d). The spectrum smoothly evolves with increasing magnetic field and the dip structures symmetric about $V=0$  move closer mimicking the smooth decrease in critical current of a superconducting point contact with increasing magnetic field.\cite{WSM17,WSM18a} This is further illustrated in Figure 1(f) where the dc current corresponding to the position of the dips (the critical current $I_{c}$) is plotted against magnetic field ($H$). Therefore, the point contact spectra presented above hint to the possible existence of a tip-induced superconducting (TISC) phase on TaAs -- similar to what was observed on the 3D Dirac semimetal Cd$_3$As$_2$\cite{WSM19,jian} and the topological crystalline insulator Pb$_{0.6}$Sn$_{0.4}$Te.\cite{WSM20} In order to further confirm the existence of a TISC, we have measured the temperature dependence of the point contact resistance (Figure 1(e)) corresponding to the spectrum presented in Figure 1(b). A resistive transition, similar to a superconducting transition, is clearly seen at $7 K$. Moreover, the transition temperature ($T_c$) systematically goes down with increasing magnetic field, as expected for a superconducting transition. A $H_{c}$-$T_{c}$ phase diagram obtained from these data is shown in Figure 1(g). However, though the above data provide sufficient hint to a TISC phase,  from the point contact data captured in a single transport regime alone does not confirm the existence of superconductivity unambiguously unless the contact can be also driven to the other regimes of transport where features related to Andreev reflection must also appear.\cite{WSM19} We have successfully explored these regimes as discussed below.

If the point contact are indeed superconducting, it should be possible to observe characterizing spectroscopic features in two other regimes of mesoscopic transport, namely the intermediate and the ballistic regime. In the intermediate regime, in addition to the critical current dominated conductance dips, two peaks symmetric about $V = 0$, which are known as hallmarks of Andreev reflection, must also appear. With further reducing the contact size it should be possible to transition to the ballistic regime where the critical current driven dips should disappear and only the double peak feature associated Andreev refection must remain.\cite{WSM19,WSM17} We have explored these two regimes and observed the expected features for supercondcuting point contacts successfully. In Figure 1(h) we illustrate a representative spectrum on TaAs in the intermediate regime showing signatures of both critical current and Andreev reflection. In Figure 1(k) we present a representative ballistic regime spectrum where only the two-peak structure symmetric about $V=0$ due to Andreev reflection survive and the dips sue to critical current disappear.\cite{BTK, WSM18} However, the magnitude of Andreev reflection is seen to be suppressed compared to what is expected for a simple elemental superconductor. The spectrum, including the conductance suppression, could be fitted well within BTK formalism, modified to include a finite transport spin polarization.\cite{BTK,Mazin,Soulen} The remarkable fit of the experimental data within the modified BTK model \cite{Anshu} is shown as a red line in the same panel. Therefore, from the observations made above it can be concluded that the point contacts on TaAs made with Ag tips are superconducting, though TaAs is not a superconductor. Hence the data presented above lead to the discovery of a new TISC phase on a Weyl semimetal TaAs.

In order to understand the nature of this new superconducting phase we concentrate on the temperature and the magnetic field dependence of the ballistic regime spectra presented in Figure 2(a) and Figure 2(b) respectively. The dotted lines show the experimental data points and the solid lines show the fits within the modified BTK model mentioned above. The remarkable match between the experimental data points and the theoretical fits must be noted here in both the panels. The gap ($\Delta$) vs. $T$ curve extracted from the data in Figure 2(a) is shown in Figure 2(c) where the expected $\Delta$ vs. $T$ curve for a BCS superconductor is also shown. The maximum superconducting energy gap is found to be 1.2 meV which drops systematically with increasing temperature and follows BCS-like temperature dependence\cite{WSM18,WSM18a} at lower temperatures, deviates slightly from BCS-like behaviour at higher temperatures closer to $T_c$ and the gap structure completely disappears at the critical temperature (7 K). The overall behavior of $\Delta$ with temperature indicates the existence of conventional $s$-wave superconductivity. However, when the deviation of the temperature dependence of $\Delta$ from the BCS behavior and the deviation of the experimentally obtained spectra for certain ballistic point contacts from the theoretical fit is considered ( as seen in Figure 2 and Figure3), the possibility of an mixed angular momentum symmetry  of the  parameter where an unconventional component is mixed with a strong $s$-wave component emerges.\cite{WSM21,WSM23,Tanaka1} Also, the possibility of the existence of multiple gaps\cite{Tanaka2} cannot be ruled out from the set of data presented here. It should be noted that for extracting spectroscopic information we have analyzed only the ballistic limit spectra where no critical current driven dips are observed and the normal state resistance remained temperature independent. 

The gap structure (double conductance peak) is seen to decrease with increasing magnetic field, as expected. A plot of $\Delta$ vs. $H$ as extracted from the fitting of the spectra in Figure 2(b) is shown in Figure 2(d). The critical magnetic field, the field at which $\Delta$ vanishes is found to be 10 kG for the point contact presented in Figure 2(b). As we mentioned before, in all the ballistic limit spectra presented above, a significant suppression of Andreev reflection was found. Since for TaAs it is known that the surface states are spin polarized,\cite{WSM_SP} the suppression of Andreev reflection can be attributed to the existence of spin polarization\cite{BTK,Soulen} at the point contacts along with superconductivity. In order to test this effect, we have modified BTK theory to include the effect of spin polarization and the modified formalism yielded remarkably good fitting of the ballistic limit point contact spectra with a large value of transport spin polarization up to 60\%. A set of three representative ballistic regime spectra showing high spin polarization are illustrated in Figure 3 (a), (b), (c) respectively. The value of the transport spin polarization is seen to decrease with increasing Z as seen in Figure 3(d). A linear extrapolation of this curve leads to an intrinsic transport spin polarization of 60\%. The measured spin polarization for a finite $Z$ is seen to increase with increasing magnetic field as shown in Figure 3 (f). The value of spin polarization  measured through this technique is lower than the 80\% spin polarization measured by ARPES\cite{WSM_SP} because point contact spectroscopy is a transport measurement and in this technique the spin polarization of the transport current is measured instead of the absolute value of the spin polarization.\cite{Mazin, Anshu}. Nevertheless, the results and the analysis presented above indicate that the super-current flowing through the TaAs point contacts is highly spin polarized.


In order to further confirm the presence of a spin polarized current along with superconductivity at TaAs point contacts, we have performed field-angle dependence of the point contact resistance where the direction of the magnetic field was rotated using a 3-axis vector magnet with respect to the direction of current. The field-angle dependent magnetoresistance data is presented in Figure 4(a). With an applied bias $V = 12 mV$ which corresponds to the normal state of the TISC, a large anisotropy in the magnetoresistance is observed which increases with increasing the strength of the magnetic field. This anisotropy in magnetoresistance can be explained if the point contact constriction is assumed to be effectively in the shape of a nano-wire and the magnetic field is rotated with respect to the direction of current flow through the nanowire. Such AMR is also seen for hybrid nanostructures involving materials having surface states with complex spin texture.\cite{AMR, AMR1} When the experiment was repeated in the superconducting state ($V = 0.25 mV$) of the TISC, the anisotropic magnetoresistance (AMR) remained equally noticeable. Therefore, the above observations conclude the co-existence of superconductivity and large spin polarization on TaAs point contacts.

In conclusion, we have shown the emergence of a tip-induced supercondcuting (TISC) phase in mesoscopic point contacts between Ag and the Weyl semimetal TaAs through transport and magneto transport measurements at various regimes of mesoscopic transport. We used a modified BTK formalism that includes spin polarization to fit the experimental data which hinted to the coexistence of superconductivity and high transport spin polarization thereby indicating the flow of spin polarized supercurrent through TaAs point contacts. The TISC, though shows the possibility of an unconventional pairing mechanism, no zero-bias conductance peak or a pseudogap could be observed in the ballistic limit data. This is in contrast to the earlier observations made on a similar TISC phase on the 3D Dirac semimetal Cd$_3$As$_2$. The surprising co-existence of superconductivity and high transport spin polarization at TaAs point contacts make the Weyl semimetals particularly interesting for spintronic applications.

\textit{Note: While preparing this manuscript we noticed that another group found some possible indication of superconductivity\cite{Wang_TaAs} in TaAs point contacts, without any direct proof\cite{Comment}.}

\textbf{Acknowledgements:} We acknowledge fruitful discussions with Professor Praveen Chaddah. R.K. thanks CSIR for junior research fellowship (JRF). G.S. would like to acknowledge partial financial support from a research grant of a Ramanujan Fellowship awarded by the Department of Science and Technology (DST), Govt. of India under grant number SR/S2/RJN-99/2011 and a research grant from DST-Nanomission under grant number SR/NM/NS-1249/2013.

\textbf{Author contributions:} The point contact experiments for exploring possible TISC in TaAs were planned and designed by GS. LA performed most of the the experiments. LA, SD, SG, RK and GS helped with the experiments and the data analysis. VS, CS and CF synthesized and characterized the high quality single crystalline TaAs samples. GS wrote the manuscript with inputs from all the co-authors.

\begin{center}
	\textbf{\underline{Figure Captions}}

\end{center}

\textbf{Figure 1: Evidence of superconductivity in thermal, intermediate and ballistic regimes of point contact transport: }(a) Schematic showing a point contact between TaAs and a sharp metal tip. (b) A representative TaAs/Ag spectrum obtained in the "thermal regime" of transport at  1.8 K. (c) Another thermal regime spectrum showing multiple conductance dips. (d) Magnetic field ($H$) dependence of the spectrum presented in (b).  (e) $H$-dependence of point contact $R-T$ showing a transition at 7 K that mimicks a superconducting transition. (f) Magnetic field dependence of the current corresponding to the dip structure ($I_c$) with $H$ extracted from the data in (d). (g) the $H_c$-$T_c$ phase diagram extracted from (e). (h) A spectrum obtained in the intermediate regime of transport showing the characteristic signatures of critical current and Andreev reflection (AR). The orange dots are experimental data points and the solid line shows a theoretical fit using modified BTK model. (k) A ballistic limit experimental spectrum showing only Andreev reflection (red dots) and an excellent theoretical fit to the spectrum (black line).

\textbf{Figure 2: Nature of superconductivity from temperature and magnetic field dependence:} (a)Temperature dependence of the ballistic limit spectra (coloured dots) and their theoretical fits (solid black lines), (b)Magnetic field dependence of the ballistic limit spectra (colored dots) and their theoretical fits (solid black lines) (c) Temperature dependence of the gap ($\Delta$). The solid line shows the expected BCS temperature dependence. (d) $H$-dependence of the gap ($\Delta$).

\textbf{Figure 3: Spin-polarization measurements:} (a), (b), (c) Three representing TaAs/Ag spectra (dots) in the ballistic limit and the corresponding theoretical fits (solid lines) showing strong transport spin polarization. (d) Barrier ($Z$) dependence of spin polarization. A linear extrapolation of this dependence to $Z = 0$ shows a large intrinsic transport spin polariation ($\sim 60\%$). (e) Magnetic field dependence of one of the ballistic point contact spectrum showing the high spin polarization. (f) Magnetic field dependence of spin polarization of the spectrum in (e).

\textbf{Figure 4: The field-angle dependence of magnetoresistance:} Anisotropic magnetoresistance of a point contact in the (a) normal state ($V = 10 mV$) and (b) in the superconducting state ($V = 0.3 mV$).

\newpage

\begin{figure}[h!]
	\includegraphics[width=.85\textwidth]{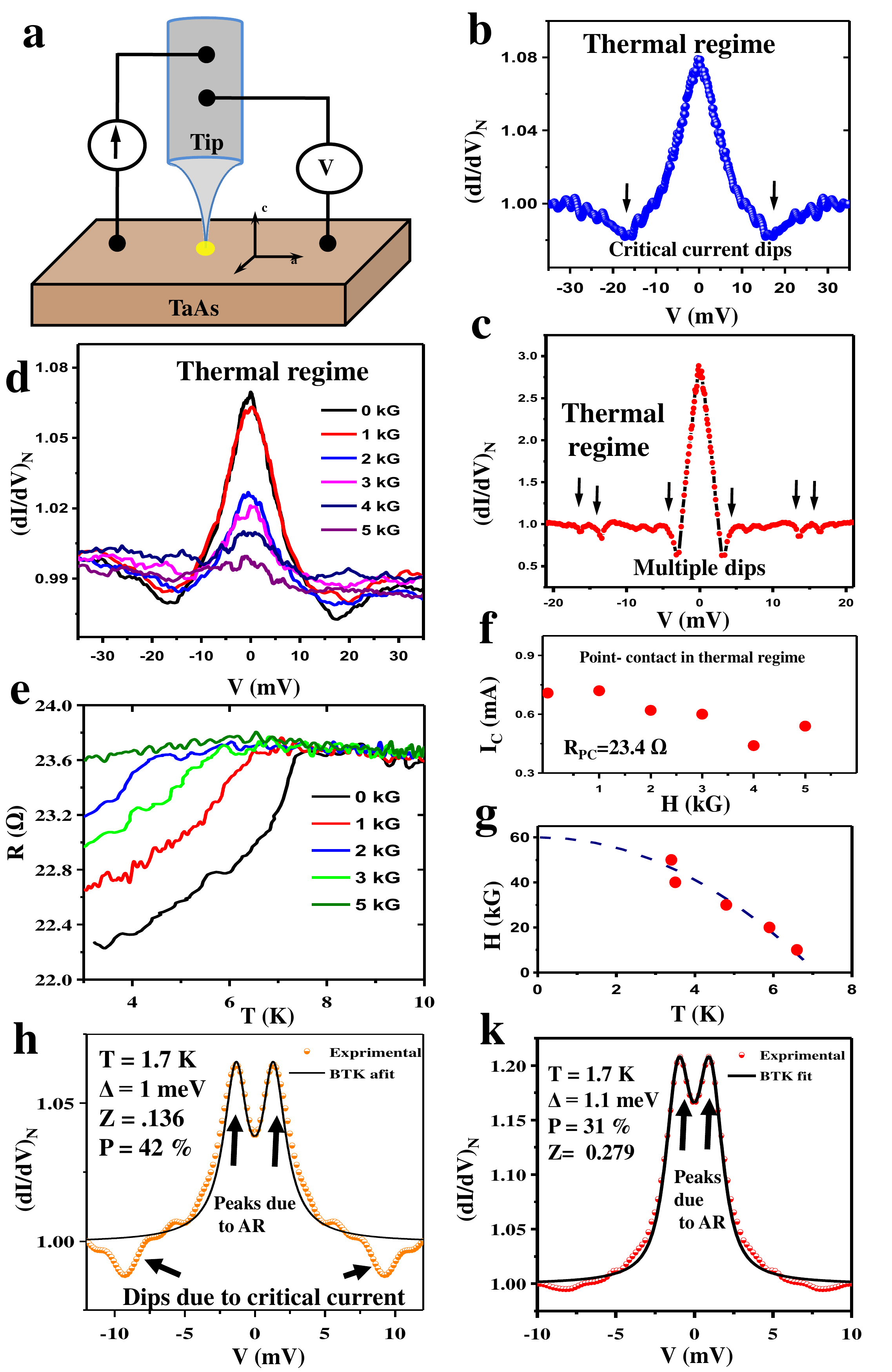}
	\label{f1}
\end{figure}

\begin{center}
	\textbf{\underline{Figure 1}}
\end{center}

\begin{figure}[h!]
	\includegraphics[width=1.1\textwidth]{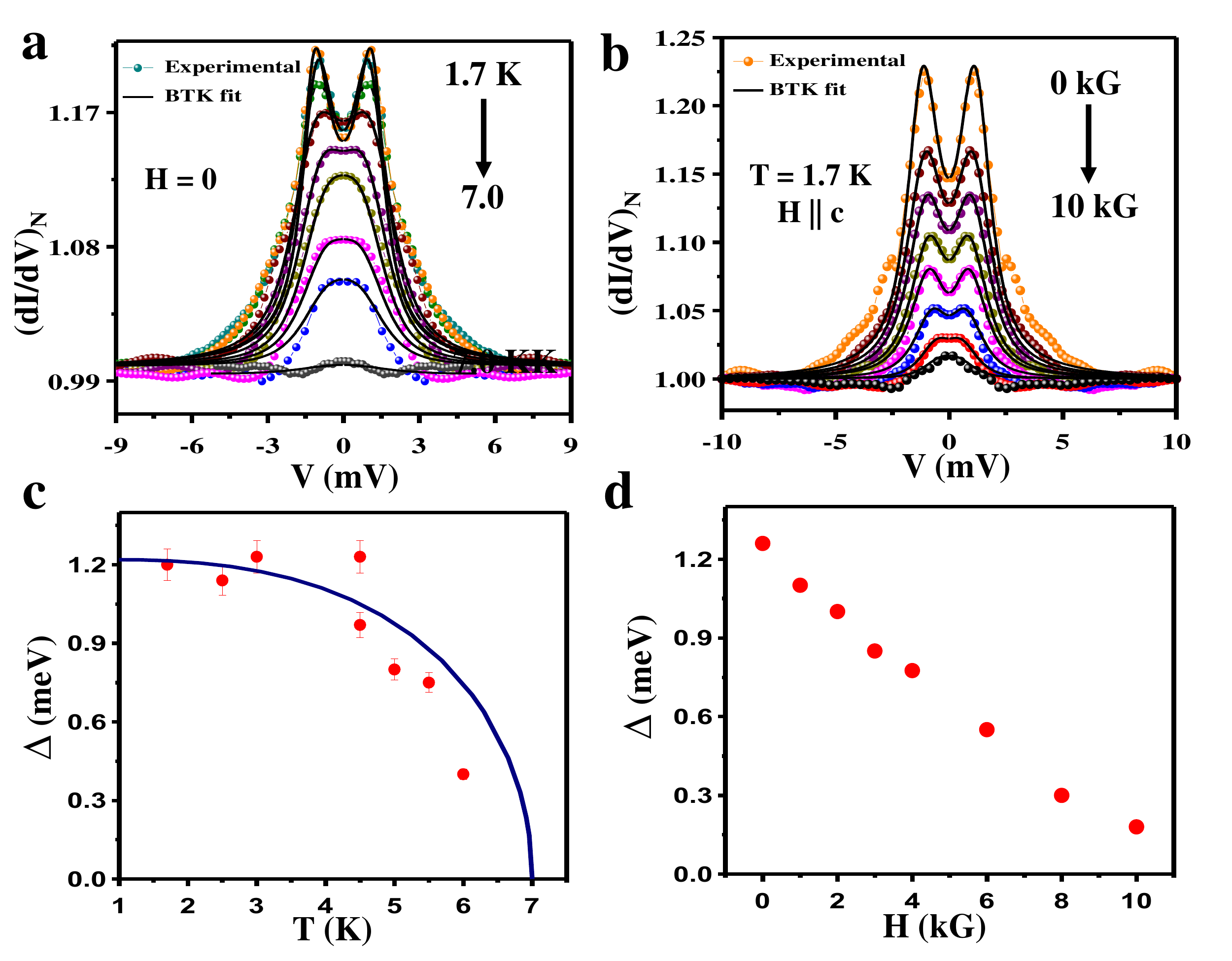}
	\label{f2}
\end{figure}

\begin{center}
	\textbf{\underline{Figure 2}}
\end{center}

\newpage

\begin{figure}[h!]
	\includegraphics[width=1.2\textwidth]{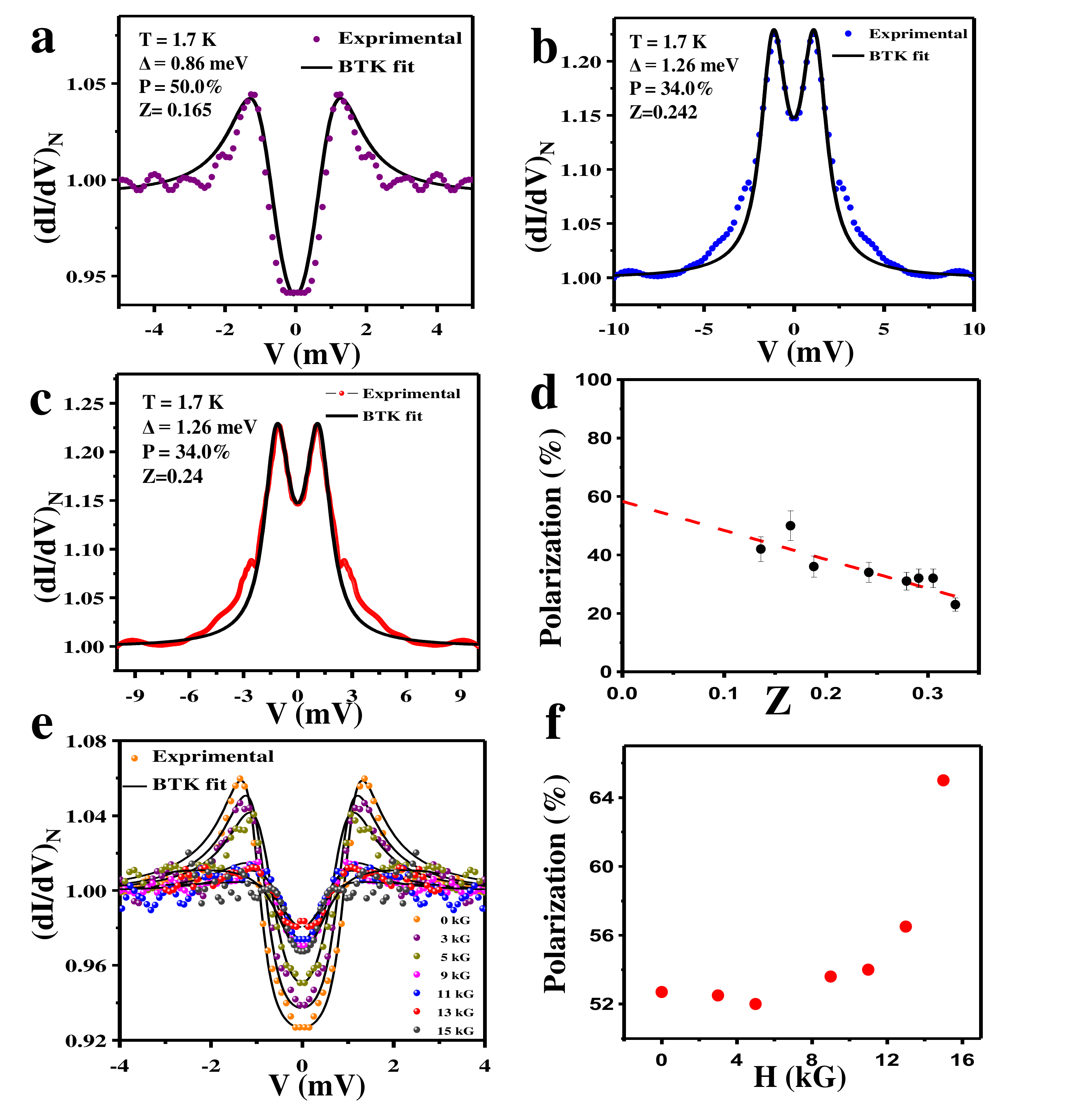}
	\label{f2}
\end{figure}

\begin{center}
	\textbf{\underline{Figure 3}}
\end{center}

\newpage 

\begin{figure}[h!]
	\includegraphics[width=1.0\textwidth]{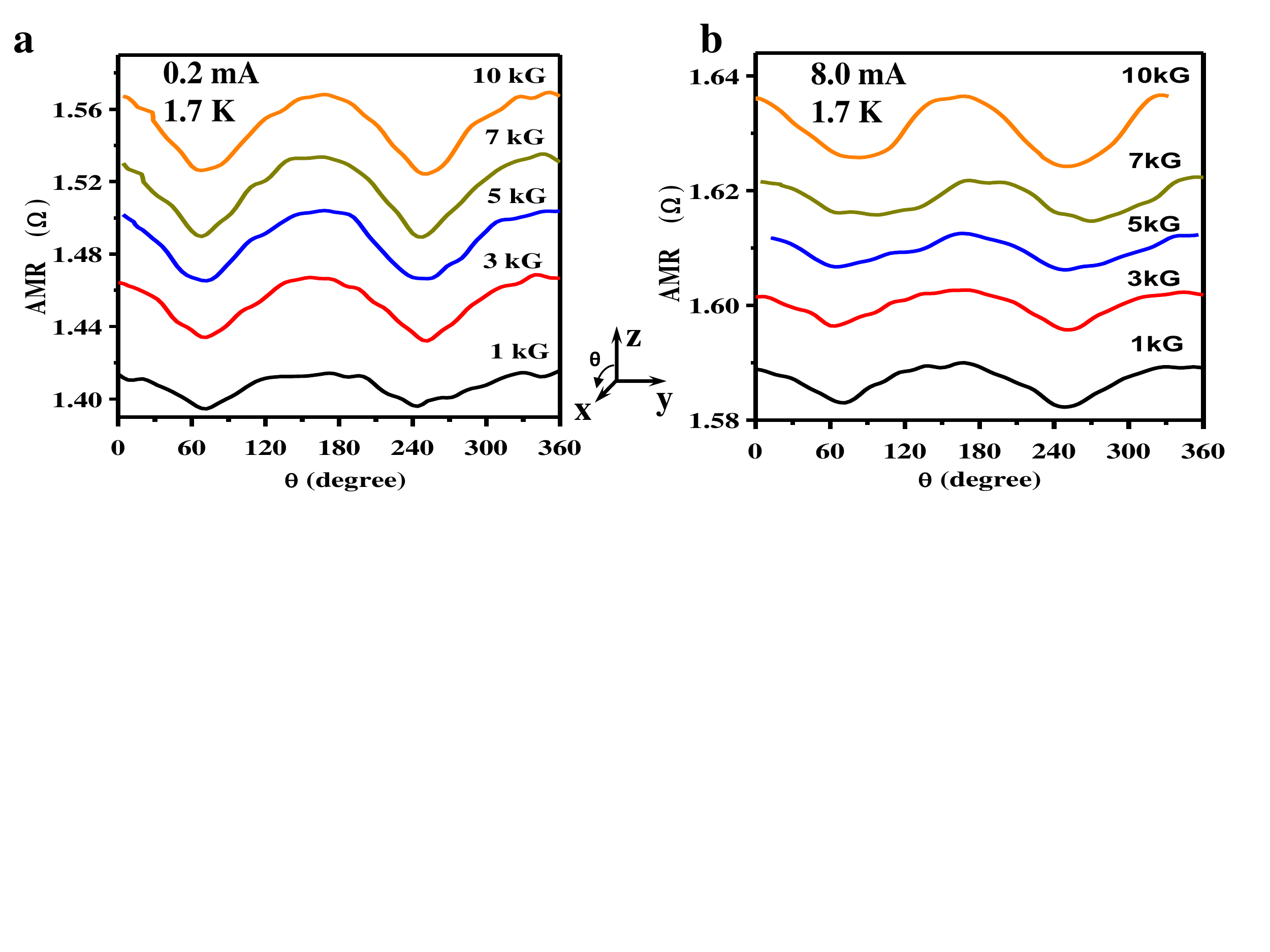}
	\label{f2}
\end{figure}

\begin{center}
	\textbf{\underline{Figure 4}}
\end{center}

\newpage

\begin{center}
	\textbf{\underline{Supplementary material}}
\end{center}

\textbf{1. Sample preparation and characterization:} All the experiments presented in this paper were carried out on high quality single crystals of TaAs. The details of sample preparation technique and characterization are provided in arXiv:1606.06649.

\textbf{2. Additional spectroscopic data in the ballistic regime:}

\begin{figure}[h!]
	\renewcommand{\figurename}{Figure S}
	\includegraphics[width=1.2\textwidth]{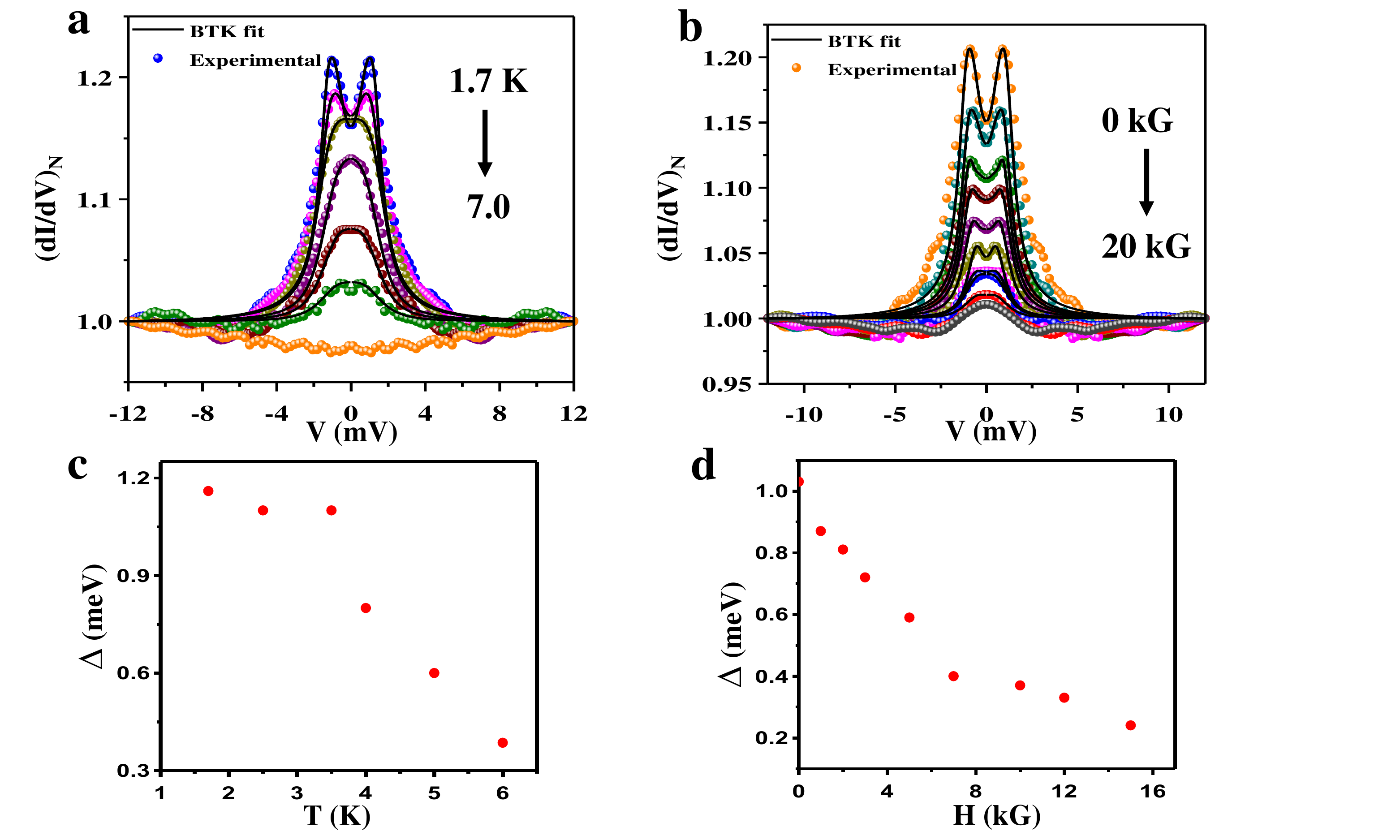}
	\caption{Magnetic field and temperature dependence of another point contact in the ballistic regime along with modified BTK fit.}
	\label{s1}
\end{figure}

\newpage

\textbf{3. Additional data used for spin-polarization measurements:}

\begin{figure}[h!]
	\renewcommand{\figurename}{Figure S}
	\includegraphics[width=0.7\textwidth]{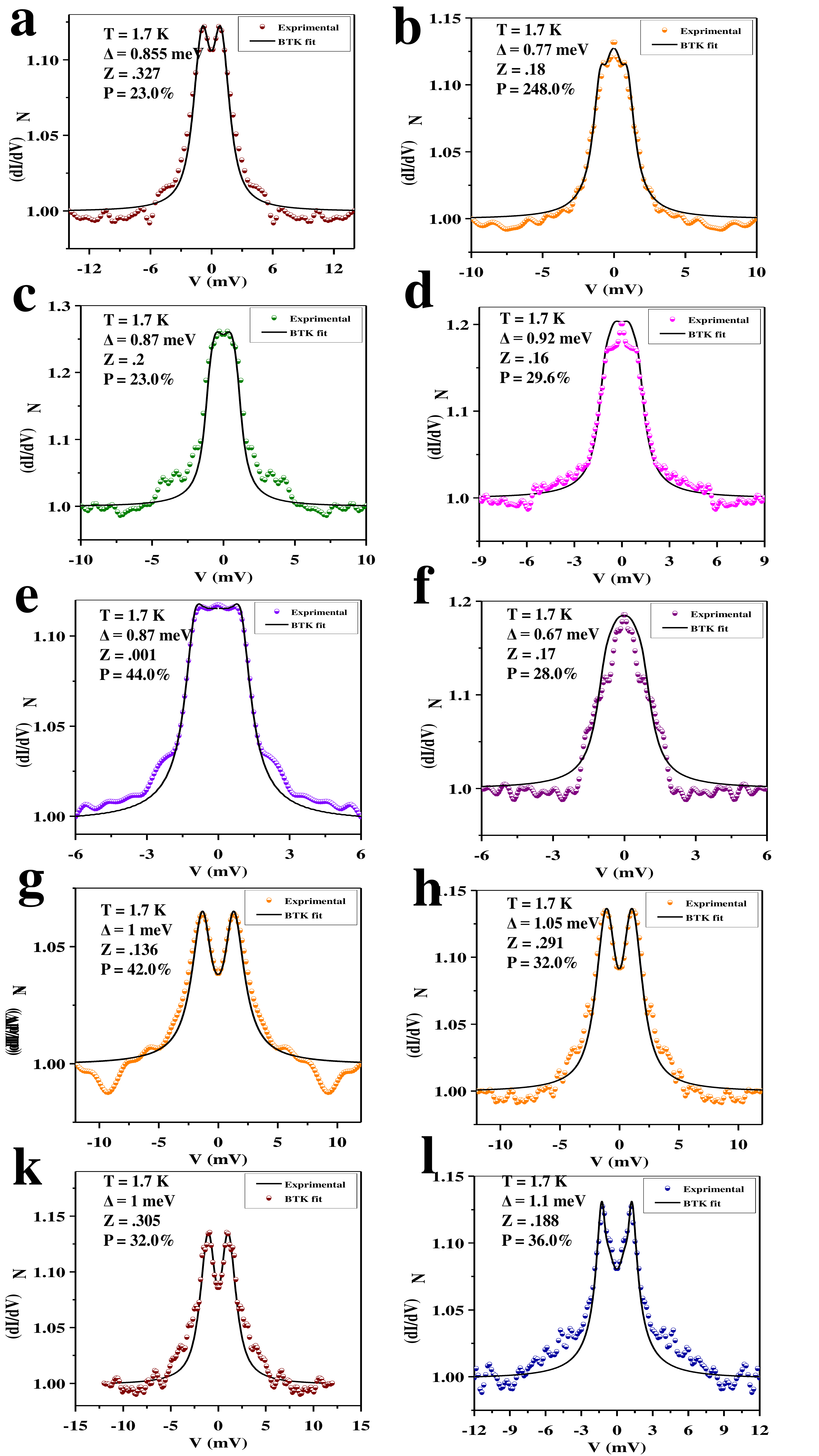}
	\caption{A collection of data and their modified BTK fit used for extracting spin polarization.}
	\label{s1}
\end{figure}

\newpage

\textbf{4. Additional data in the thermal regime:}
\begin{figure}[h!]
	\renewcommand{\figurename}{Figure S}
	\includegraphics[width=1.2\textwidth]{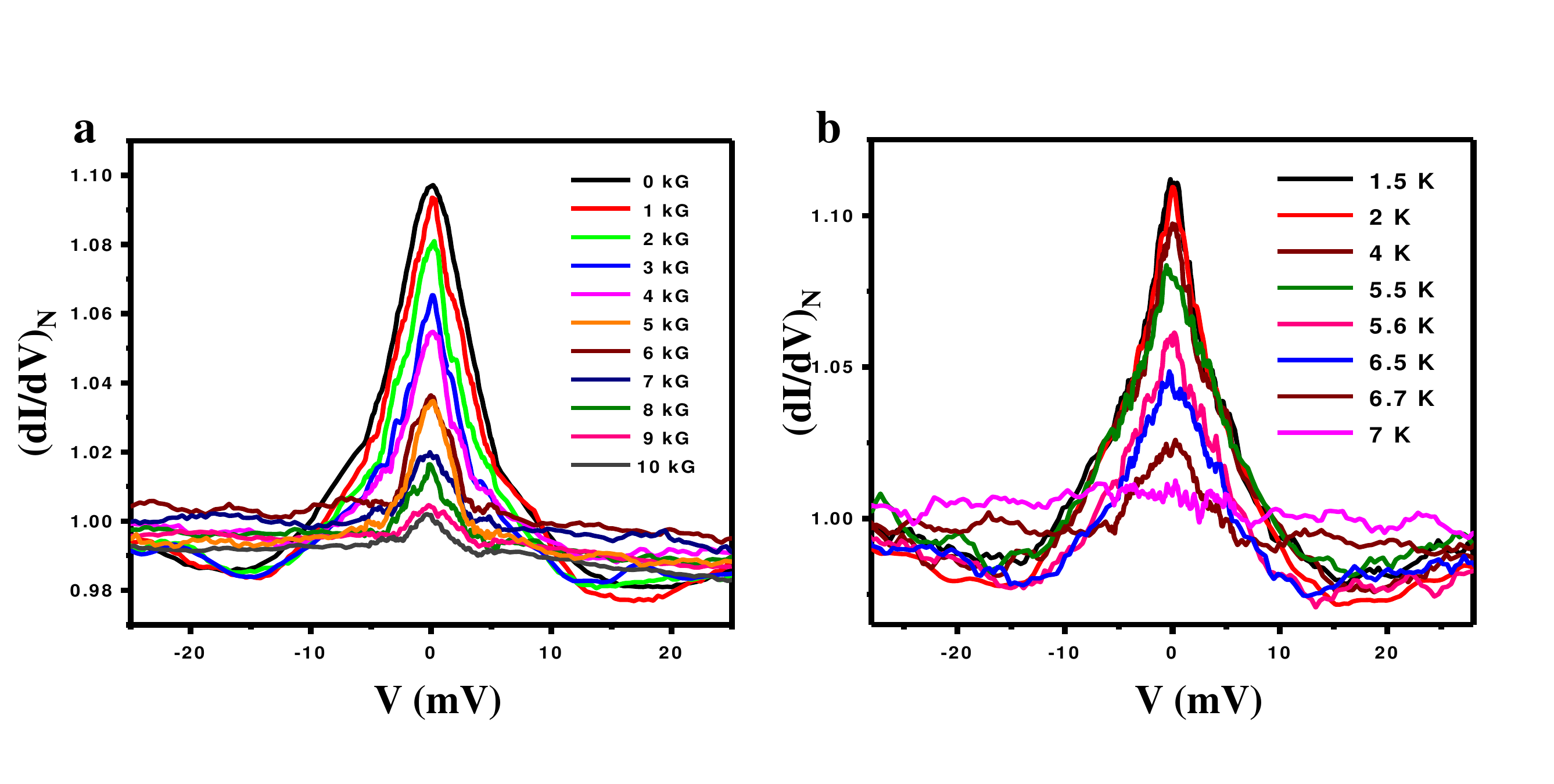}
	\caption{Magnetic field and temperature dependence of another point contact in the thermal regime.}
	\label{s1}
\end{figure}

\newpage

\textbf{5. A collection of different types of contact-geometry dependent data obtained in the thermal regime:}
\begin{figure}[h!]
	\renewcommand{\figurename}{Figure S}
	\includegraphics[width=1.2\textwidth]{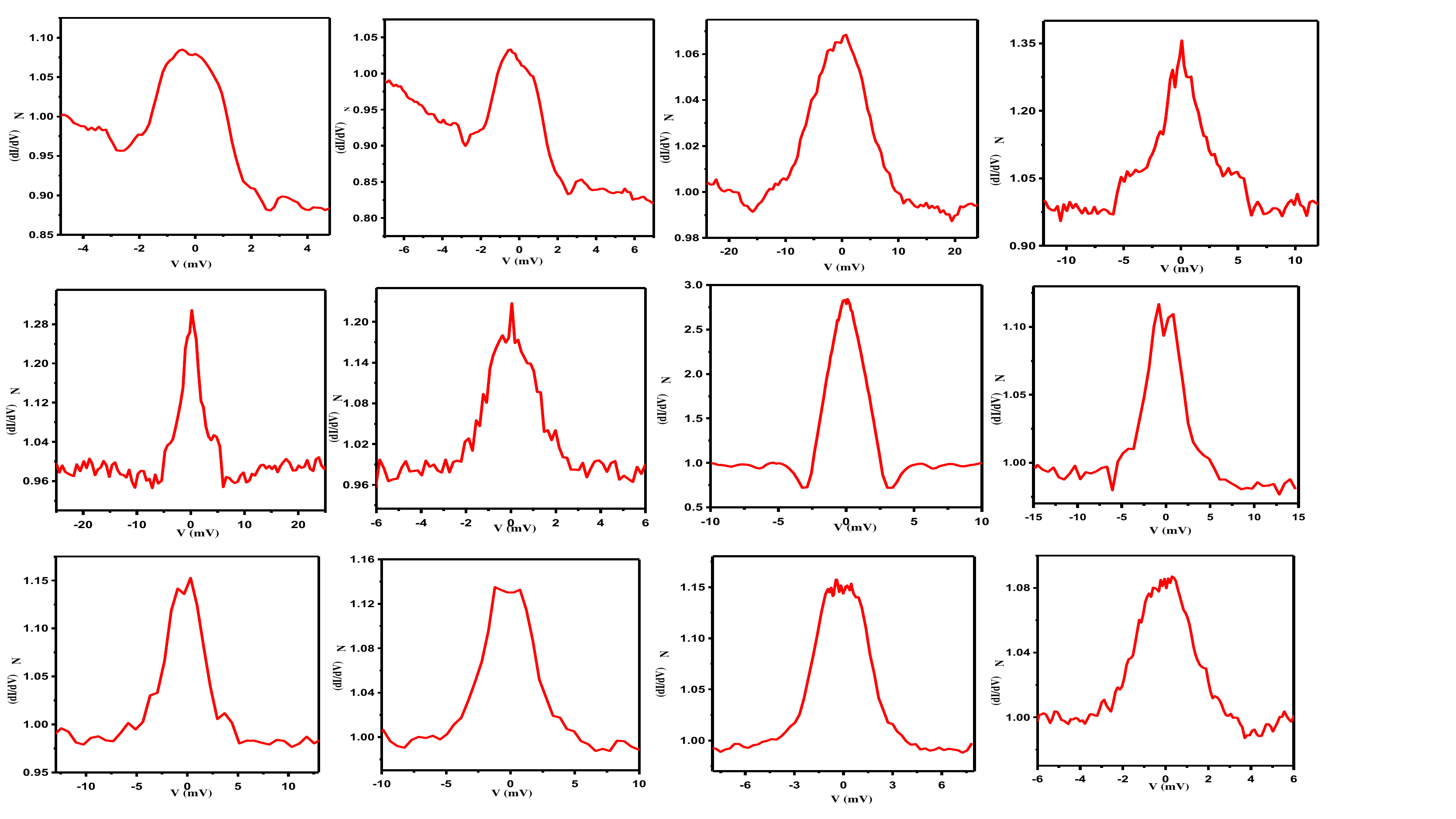}
	\caption{A collection of thermal limit point-contacts showing geometry dependence of the point contacts.}
	\label{s1}
\end{figure}

\end{document}